\documentclass[aps,prl,reprint,groupedaddress]{revtex4-1}
\usepackage{graphicx}
\usepackage[T1]{fontenc}       
\usepackage[utf8]{inputenc}

\begin{document}

\title{Interlayer Charge Transfer and Defect Creation in Type I van der Waals Heterostructures}

\author{G. Nayak}
\author{S. Lisi}
\author{W-L. Liu}
\affiliation{Univ. Grenoble Alpes, CNRS, Grenoble INP, Institut Néel, 38000 Grenoble, France}
\author{T. Jakubczyk}

\altaffiliation{Present address:  Department of Physics, University of Basel, 4056 Basel, Switzerland}
\author{P. Stepanov}
\author{F. Donatini}
\affiliation{Univ. Grenoble Alpes, CNRS, Grenoble INP, Institut Néel, 38000 Grenoble, France}
\author{K. Watanabe}

\author{T. Taniguchi}
\affiliation{National Institute for Materials Science, Tsukuba, 305-0044, Japan}
\author{A. Bid}
\affiliation{Department of Physics, Indian Institute of Science, Bangalore 560012, India}
\author{J. Kasprzak}
\author{M. Richard}
\author{V. Bouchiat}
\author{J. Coraux}
\author{L. Marty}
\author{N. Bendiab}
\author{J. Renard}
\email{julien.renard@neel.cnrs.fr}
\affiliation{Univ. Grenoble Alpes, CNRS, Grenoble INP, Institut Néel, 38000 Grenoble, France}

\date{\today}

\begin{abstract}
Van der Waals heterostructures give access to a wide variety of new phenomena that emerge thanks to the combination of properties brought in by the constituent layered materials.  We show here that owing to an enhanced interaction cross section with electrons in a type I van der Waals heterostructure, made of single layer molybdenum disulphide and thin boron nitride films, electrons and holes created in boron nitride can be transferred to the dichalcogenide where they form electron-hole pairs yielding luminescence. This cathodoluminescence can be mapped with a spatial resolution far exceeding what can be achieved in a typical photoluminescence experiment, and is highly valuable to understand the optoelectronic properties at the nanometer scale. We find that in heterostructures prepared following the mainstream dry transfer technique, cathodoluminescence is  locally extinguished, and we show that this extinction is associated with the formation of defects, that are detected in Raman spectroscopy and photoluminescence. We establish that to avoid defect formation induced by low-energy electron beams and to ensure efficient transfer of electrons and holes at the interface between the layers, flat and uniform interlayer interfaces are needed, that are free of trapped species, airborne ones or contaminants associated with sample preparation. We show that heterostructure fabrication using a pick-up technique leads to superior, intimate interlayer contacts associated with significantly more homogeneous cathodoluminescence. 
\end{abstract}

\maketitle

\section{Introduction}

The development of deterministic stacking of individual or few layers from layered materials such as graphite, boron nitride, or transition metal dichalcogenides (TMDCs) in the last few years allows now to build van der Waals heterostructures at will \cite{Novoselov2016}. Even though such structures are produced in conditions that are far from ultra high vacuum conditions usually required to obtain very high quality heterostructures, careful preparation yields clean enough interfaces to observe phenomena requiring efficient electronic coupling between layers. For instance, long-lived interlayer excitons were observed \cite{Rivera2015,Rivera2016}, interlayer exciton-phonon coupling was reported in WSe$_2$/hexagonal boron nitride (h-BN) \cite{Jin2016}, interlayer phonon coupling was demonstrated in MoS$_2$/graphene \cite{Li2017} as well as in TMDC-based van der Waals heterostructures \cite{Lui2015}. These achievements demonstrate that both inter-layer electronic and structural coupling can be obtained in such heterostructures. Moiré superlattices, which are the van der Waals (soft) counterpart of dislocation networks in heteroepitaxial three-dimensional semiconductors, enrich the electronic and optical properties \cite{Kang2013,Wu2017,Pan2018,Huang2018,Tran2018}.

\begin{figure*}
\includegraphics{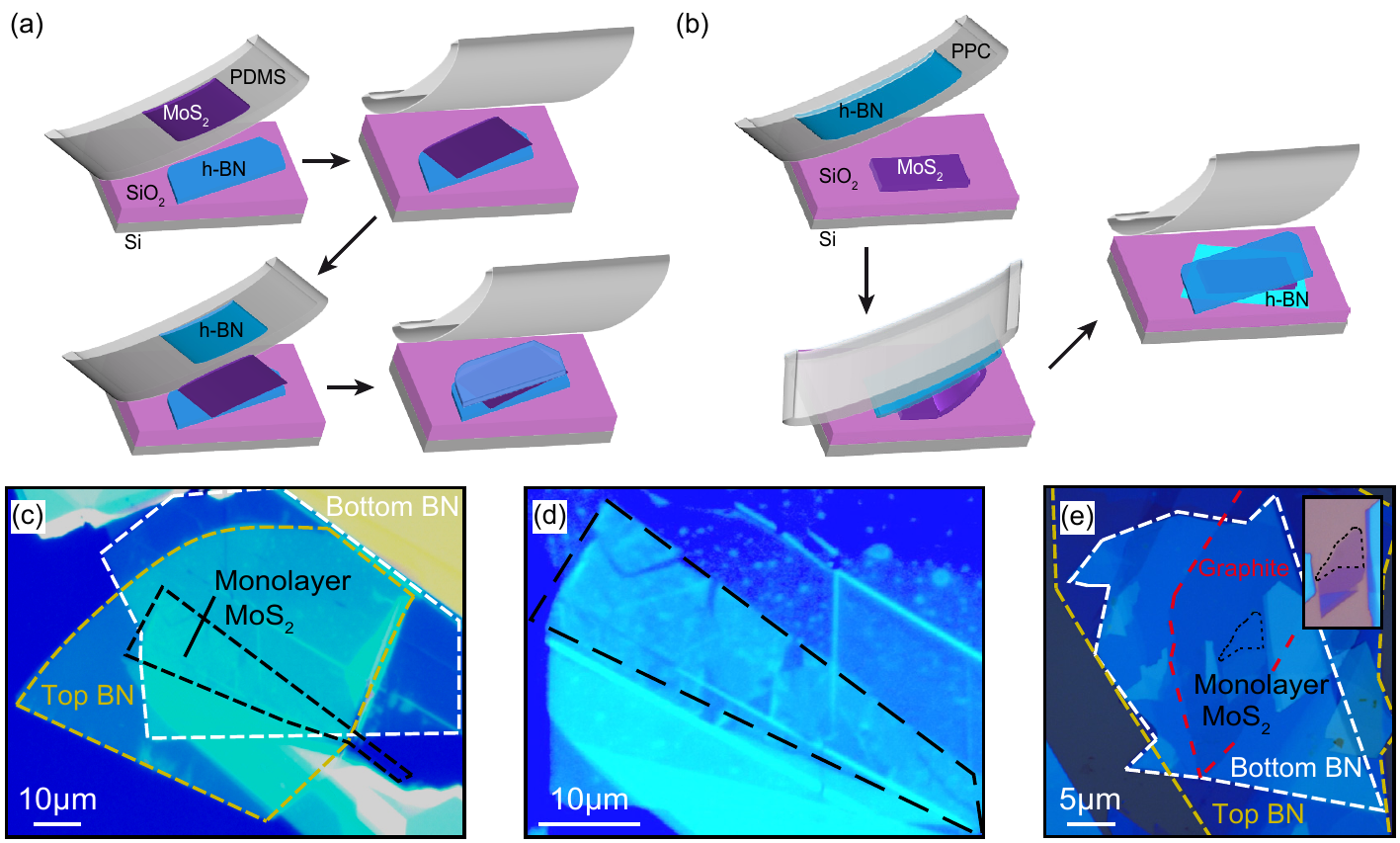}
  \caption{\textbf{Samples preparation and optical images.} Sketches of the two preparation methods used in this work: PDMS stamping (a) and PPC pick-up (b). Optical micrographs of the full stack prepared with PDMS stamping (c). A micrograph focusing on the monolayer MoS$_2$ regions is shown in (d). The sample prepared using the pick-up technique is shown in (e). The inset presents the region of interest of MoS$_2$ prior to pickup, i.e. on SiO$_2$, in which case the optical contrast is higher.}
  \label{sample}
\end{figure*}

Direct observations with high resolution transmission electron microscopy indeed revealed locally perfect crystalline interfaces between two-dimensional materials, with no apparent defects and only a van der Waals gap a few {\AA}ngstr\"om-thick, devoid of impurity species \cite{Rooney2017}. Even though beam-induced damages can be reduced by lowering the energy of the electron beam to a few 10~keV in the transmission electron microscope column \cite{komsa2012}, such analysis is destructive (the samples need to be cut and thinned down to a few nanometer), and it is tedious to extend it at the scale of the entire heterostructure. On the contrary optical hyperspectral microscopies, mapping excitonic (photoluminescence) or vibrational (Raman) interlayer modes of the heterostructures, require no additional sample preparation, and provide indirect information on the quality of interfaces \cite{Lui2015,Alexeev2017}. Their downside is their limited spatial resolution, which conceals information on, \textit{e.g.}, strain or electronic doping variations at scales below a few 100~nm.

Cathodoluminescence (CL) is a powerful tool to study opto-electronic properties at the nanometer scale, when implemented in a scanning electron microscope. Here, an electron beam of adjustable energy in the keV range excites electrons and holes, that can form electron-hole pairs (excitons) and recombine radiatively, giving local spectroscopic information. In that case spatial resolution is linked to the size of the excitation source, \textit{i.e.} the electron beam which is as small as a few nanometers, rather than limited by the optical diffraction limit. In fact, spatial resolution in such experiment could ultimately be set by the diffusion of the excitons. It can reach several hundreds of nanometers at room temperature in monolayer TMDCs but is expected to be strongly quenched at low temperature due to the enhancement of the radiative recombination rate \cite{Kulig2018}.

Due to the low interaction with the electron beam, atomically thin layers are expected to produce a small signal below the detection limit of most instruments. In fact, no CL could be measured so far on a free-standing TMDC single layer. Van der Waals heterostructures can be used to artificially enhance the interaction by encapsulating the active layer into an electronic barrier, very much like semiconductor quantum wells are built: a small band gap material (well) is surrounded by a larger band gap material (barrier). Electron beam irradiation generates hot electrons and holes inside the barrier, which can be arbitrarily thick and hence produce a significant population of hot charge carriers that relax and can be transferred into the low band gap material. This approach has been recently demonstrated with h-BN as a barrier and a single layer TMDC (MoS$_2$,WS$_2$, WSe$_2$) as the active lay and the decisive effect of h-BN capping was demonstrated by varying its thickness \cite{zheng17} .

So far cathodoluminescence was observed only in limited area of van der Waals heterostructures, and the absence of cathodoluminescence was ascribed to a locally poor contact between the h-BN and TMDC surfaces \cite{zheng17}. In this scenario, it is implied that in that case charge carrier transfer between the barrier and active material is inefficient, and hence no exciton can be formed in the latter material. Poor-contact-regions are indeed very common in heterostructures. They correspond to blisters trapped at the interfaces, where contaminants associated with the manipulation of the two-dimensional materials gather \cite{Pizzocchero2016}.

\begin{figure*}
\includegraphics{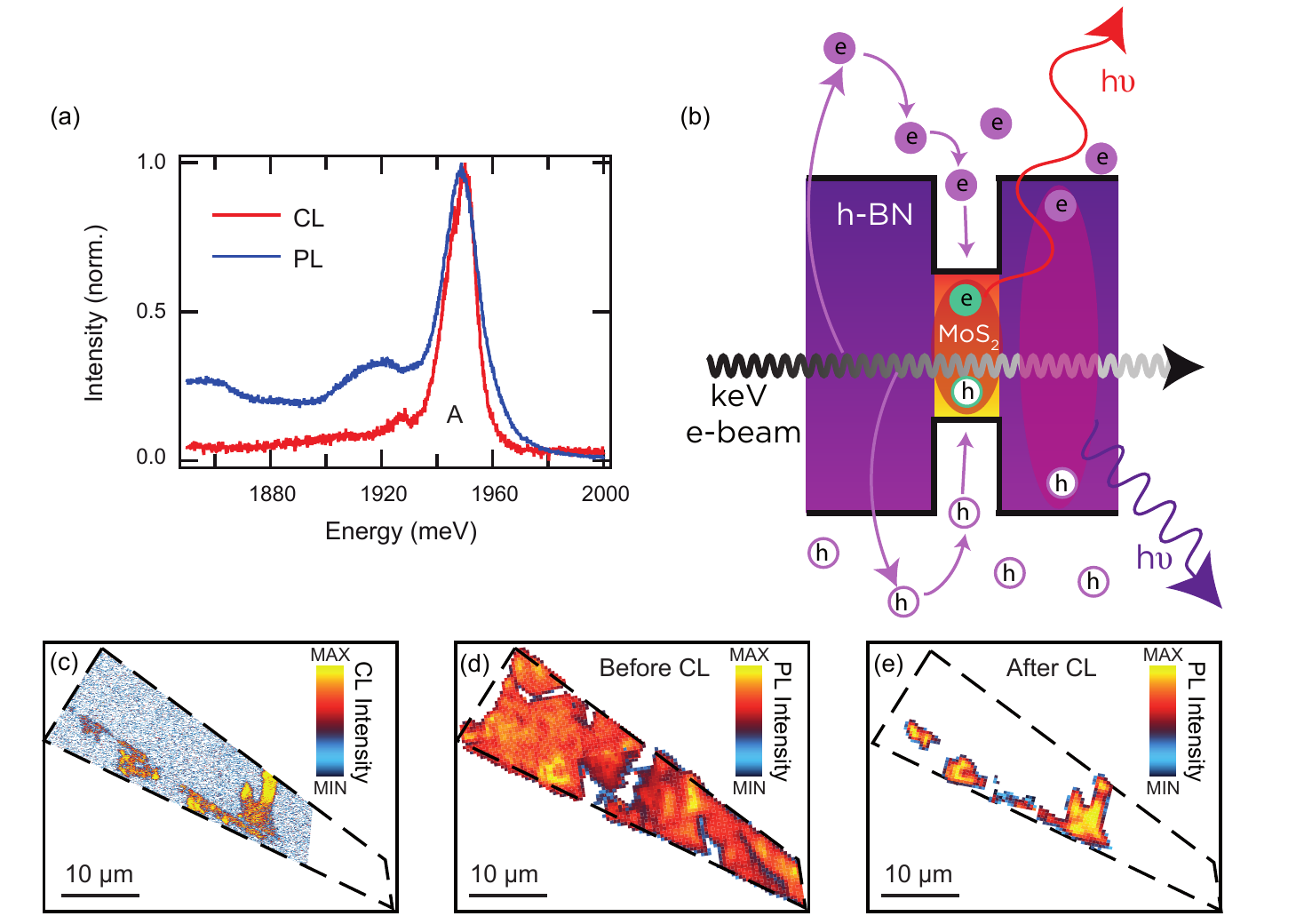}
  \caption{\textbf{Optical characterization using Photoluminescence (PL) and Cathodoluminescence (CL).} A comparison of spatially-resolved spectra at 5~K obtained in PL and CL is presented in (a). The spectra are acquired within less than 1~$\mu$m of each other. In both cases, the signal is dominated by neutral A-exciton emission. The process of luminescence excitation in a van der Waals heterostructure with an electron beam is presented in (b). The vertical axis represents energy (band diagram).
	The spatially-resolved integrated CL intensity of the exciton is shown in (c) showing some strong inhomogeneities. This is not the case for the integrated PL intensity measured before CL at room temperature (d). In contrast, integrated PL intensity measured after CL (e) shows a strong inhomogeneity presenting a spatial correlation with the CL mapping.}
  \label{CL}
\end{figure*}

Here we demonstrate that indeed an intimate contact between the materials is key to observe efficient cathodoluminescence. Conversely, we find that in presence of contaminants at the interfaces cathodoluminescence is quenched. This quenching is also observed in photoluminesence performed after the cathodoluminescence. It is hence not only the signature of an inefficient transfer of electrons and holes, as it was thought, but it proves the creation of crystal defects inside MoS$_2$ that presumably strongly promote non-radiative exciton recombination. Such defects are detected in Raman spectroscopy. We trace back the origin of defects to a possible chemical reaction between trapped species and the pristine MoS$_2$, promoted by the electron beam. We show that the spatial uniformity of the cathodoluminescence response of the heterostructures can be greatly enhanced by reducing the amount of contaminants at interfaces (in particular, the amount of trapped blister), using polypropylene carbonate (PPC) to pick-up and release the different materials of the heterostructure \cite{Pizzocchero2016}.

\section{Experiment}
\subsection{Van der Waals heterostructure preparation}
The van der Waals heterostructures studied in this work were assembled by two methods (Figure~\ref{sample}a,b). A dry viscoelastic transfer method using polydimethylsiloxane (PDMS) \cite{Castellanos2014}, involving two successive stamping, of MoS$_2$ and h-BN, and a pick-up technique using PPC \cite{Wang2013,Pizzocchero2016}, allowing to directly position a h-BN/MoS$_2$ stack, were both employed to encapsulate MoS$_2$ single-layers between h-BN layers ($\sim$ 20~nm-thick). The details are provided in \cite{suppinfo}. The optical pictographs of the h-BN/MoS$_2$/h-BN heterostructure are displayed in Figures~\ref{sample}c,d,e. The MoS$_2$ regions encapsulated between the top and bottom h-BN layers extend across tens of micrometers. We will first focus on the sample prepared using dry viscoelastic stamping (Figure~\ref{sample}a,c,d).

\begin{figure*}
\includegraphics{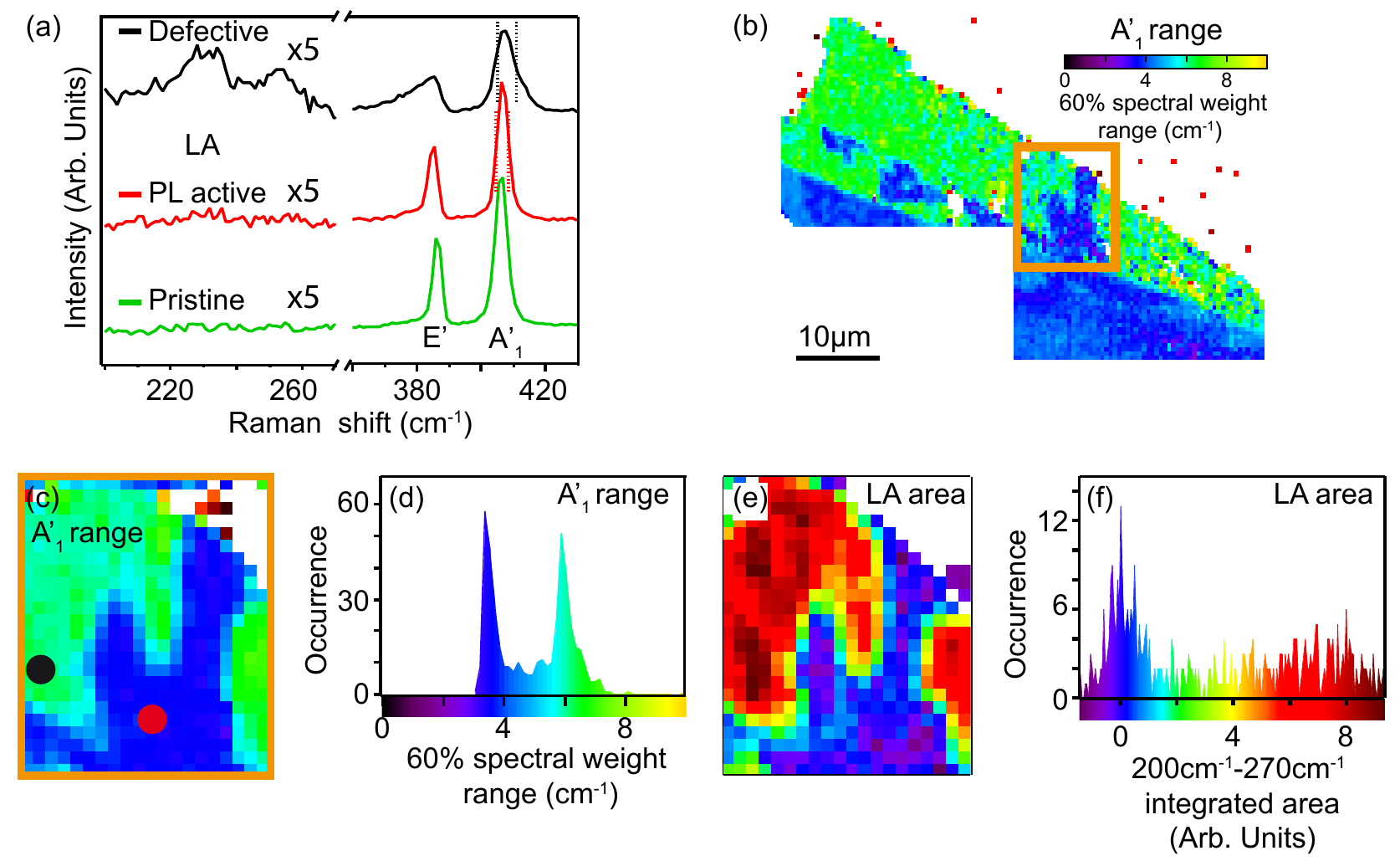}
  \caption{\textbf{Raman spectroscopy and spatial mapping of defects.} Representative spectra of the different regions are presented in (a): pristine (i.e. before CL), PL active and PL inactive/defective. In addition to a broader A$^{'}_{1}$ peak, the defective regions show the emergence of Raman signal around 230~cm$^{-1}$, which we refer to as the LA area. Such signal in this area is also a signature of defects. The spatially-resolved width of the A$^{'}_{1}$ peak is used as a metric for the presence of defects created during the CL experiment, as discussed in the main text. The map shown in (b) presents a strong spatial correlation with both CL and PL maps measured after CL (see Figure~\ref{CL}~c,e). A map with a better spatial resolution of a smaller area is presented in (c). The black and red dots in (c) correspond to the defective and PL active spectra shown in (a), respectively. The histogram of A$^{'}_{1}$ widths is presented in (d), in which two peaks attributed to the absence (blue, small width) or presence (green, large width) of defects appear. The color scale meaning in (c) is provided in (d). Note that the exact same color scale is used in (b), (c) and (d) for comparison. (e) shows the spatially-resolved integrated intensity in the LA area (see (a)) at the same position as in (c). The color scale used in (e) is detailed in (f), which shows also the histogram of integrated area for the LA region.}
  \label{Raman}
\end{figure*}

\subsection{Spatially non-uniform cathodoluminescence}
Figure~\ref{CL}a  shows a typical CL spectrum of single-layer MoS$_2$ encapsulated in h-BN with the PDMS transfer technique. A strong emission is centered around 1950 meV corresponding to the energy of the recombination of the neutral A-exciton (EX) in MoS$_2$. The high energy of the electrons (1~keV) compared to the h-BN band gap allows to excite electrons and holes directly in the h-BN, as witnessed by near band edge luminescence from h-BN \cite{suppinfo}. Due to its thickness (top: 18~nm, bottom: 22~nm) a significant number of electrons and holes can be generated in h-BN. Using Monte-Carlo simulations, we have shown that at 1~keV, the absorption length in h-BN is of the order of several tens of nm \cite{suppinfo}.  When the contact with single-layer MoS$_2$ is intimate, electrons and holes can be transferred into the latter material which has a much smaller band gap than h-BN. They can then form an EX that will eventually contribute to luminescence when recombining radiatively (A-exciton emission in Figure~\ref{CL}a). Alternatively EX can be formed directly in h-BN and either recombine radiatively or relax into MoS$_2$. The full CL process is illustrated in Figure~\ref{CL}b.

For comparison with CL, a PL spectrum taken in the same area is presented in Figure~\ref{CL}a. We see that the spectra are very similar proving their common origin, that is radiative recombination (photon emission) of the A exciton of single layer MoS$_2$. We also want to stress that linewidths are both of the order of 10 meV. It shows that no additional broadening is brought by using a high energy electron beam (in the CL experiment) as the excitation instead of light (PL experiment). This is a very important point, and rather unexpected when referring to literature on CL. Large broadenings, associated with the large number of free charges generated, can indeed often be observed \cite{Rodriguez1997}. The lack of broadening demonstrated here is a strong asset for the potential application of CL in mapping properties of van der Waals heterostructures at the nanoscale. We note that the excitation energy used in our PL experiment is not sufficient (unlike in the CL experiment) to excite electrons and holes directly in h-BN, so we do not observe luminescence from this material in these conditions.

The spatial mapping of the CL intensity measured at 1937 meV (+/- 30 meV ) is presented in Figure~\ref{CL}c. Bright and dark regions are observed, in accordance with a previous report \cite{zheng17}. In area of the sample where MoS$_2$ is not encapsulated with h-BN, no CL is detected; CL is only detected in localized regions of the h-BN-encapsulated MoS$_2$ \cite{suppinfo}. Encapsulation hence appears necessary, because h-BN is the source of electrons and holes that will eventually recombine in MoS$_2$, but not sufficient.
An absence of CL in presence of h-BN is in principle surprising, and this was proposed as an evidence of a poor contact between h-BN and MoS$_2$ \cite{zheng17}. Atomic force microscopy (AFM) reveals that the corresponding regions show a significant roughness and bubble-like features at the surface \cite{suppinfo}. We relate these observations to the presence of blisters which are filled with species (airborne, contaminants from the polymer stamp) and are trapped at the interface between h-BN and MoS$_2$. The regions exhibiting CL actually appear very flat in AFM \cite{suppinfo}, with a root mean square roughness in the range of 2~\AA, typically several times lower than for other regions. This is indicative of very smooth and flat (buried) interfaces between the materials.

Prior to the measurement of CL maps, we measured PL maps. Figure~\ref{CL}d displays the integrated PL intensity of the EX (sum of neutral and charged excitons contributions, the latter appearing as a low-energy shoulder ). Unlike for CL, the MoS$_2$ layer appears essentially bright, except at locations where it is physically cracked. There are variations of the PL intensity and position which are attributed to local changes (strain, doping, coupling to h-BN)  but no quenching. A straightforward interpretation for the only partial spatial correlations between CL and PL maps on one hand, and the clear correlation between dark regions as found with CL and rough regions as found in AFM on the other hand, relates to the effectiveness of electrons and holes transfer at the interface between MoS$_2$ and h-BN \cite{zheng17}. Indeed, it seems reasonable to expect that the presence of species intercalated in between the two materials, in the form of blisters or in other forms, hinders charge transfers. However, we will now see that other effects prevail.

\subsection{Quenching of luminescence by defects.}
While before exposure to the electron beam (used for the CL measurement), PL revealed essentially bright regions, the PL map  measured after CL strongly correlates with the CL map, showing the same dark regions (compare Figures~\ref{CL}c,e). It appears thus that the irradiation by the electron beam locally quenches luminescence, regardless of the source of excitation used to observe it. This observation questions the common conception that CL is quenched only by charge transfer hindrance at interfaces; instead it suggests a more invasive effect of the electron beam, damaging MoS$_2$.

To address this possibility, we analyzed the vibrational properties of MoS$_2$, before and after electron beam irradiation. We performed Raman spectroscopy measurements at room temperature using an excitation wavelength of 532~nm (see \cite{suppinfo}). In the regions showing CL, we always observe Raman spectra with two prominent peaks (Figure~\ref{Raman}a), corresponding to the intralayer in-plane E$^{'}$ and out-of-plane A$^{'}_{1}$ modes \cite{lee2010,molina2011}. These modes correspond to first-order Raman processes and arise from single phonon at the center of the Brillouin zone. Small deviations from single Lorentzian lineshapes, expected for such processes, can nevertheless be extracted from the experimental spectra of pristine monolayer MoS$_2$ \cite{suppinfo}. Such deviations likely come from the additional contributions of doubly resonant Raman (DRR) processes. For the E$^{'}$ mode, a low energy shoulder  was already reported in pristine monolayer MoS$_2$ \cite{Mignuzzi2015,Carvalho2017}. As the E$^{'}$ mode is doubly degenerate, such shoulder could be attributed to a degeneracy lifting induced by strain or doping \cite{molina2011}. 

Electron beam irradiation has no effect on the spectra in regions showing CL. In regions showing no CL, on the contrary, electron irradiation has strong effects. Prior to irradiation, we also exclusively observe signatures of the E$^{'}$ and A$^{'}_{1}$ modes. After irradiation, the corresponding peaks appear broader and show some structure, and new modes are found at lower frequency, around 230~cm$^{-1}$ (Figure~\ref{Raman}a). They form a band, that is referred to as LA(M) in the literature \cite{frey1999,Mignuzzi2015,Carvalho2017}. While an exact fitting procedure would require a full theoretical description of Raman intensities including DRR contributions, such complex spectra have been fitted with a sum of Lorentzian in the literature \cite{Livneh2015,Mignuzzi2015,Carvalho2017}. Within this approximation, we can associate a peak with a given phonon in the band structure. The band around 230~cm$^{-1}$ was shown to be defect-induced. A detailed, excitation energy dependent study, showed that this band arises from DRR processes involving one phonon; elastic scattering by a defect ensuring momentum conservation \cite{Carvalho2017}. This band has three main contributions: a van Hove singularity in the phonon density of states between the K and M points, the LA branch in vicinity of the M point (LA(M)) and the LA branch in the vicinity of the K point (LA(K)) \cite{Carvalho2017}. In addition we observe another defect-activated contribution near 250~cm$^{-1}$ which was also reported by Mignuzzi et al. \cite{Mignuzzi2015} but left unassigned.

We now discuss the evolution of the spectra in the vicinity of the E$^{'}$ and A$^{'}_{1}$ modes upon electron irradiation in regions showing no CL. Defect-induced contributions are visible close to both E$^{'}$ and A$^{'}_{1}$. They have been attributed to phonons in the vicinity of the M points on the TO, LO and ZO branches \cite{Mignuzzi2015}. We have quantitatively analyzed the weight of those defect-induced contributions in order to spatially map the occurrence of defects. The details of the analysis are presented in \cite{suppinfo}. In Figure ~\ref{Raman}b-d , we image the presence of defects in the sample using this approach. We see that there is a remarkable spatial correlation between defect mapping using Raman spectroscopy and CL/PL mapping (Figure ~\ref{CL}c,e). In addition, to strengthen the validity of our method, we perform a similar defect-mapping analysis based on the LA(M) band (Figure ~\ref{Raman}e,f). We see here that regions with a measurable contribution around 230~cm$^{-1}$ are the ones that are defective and thus CL/PL inactive. The quenching of luminescence by electron beam irradiation is hence related to defect creation in MoS$_2$.

\subsection{Strong luminescence in case of clean interfaces}
We now turn our attention to the origin of defects creation and the possibility to locate regions with an intimate contact between h-BN and MoS$_2$ prior to CL and the creation of defects. 

\begin{figure}
\includegraphics{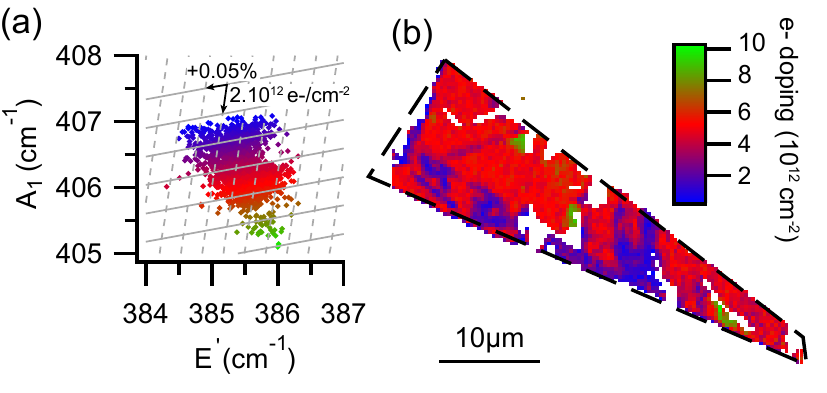}
  \caption{\textbf{Spatially resolved interlayer contact mapping through doping analysis before cathodoluminescence}. In (a) the energy shifts of both the E$^{'}$ and A$^{'}_{1}$ Raman modes acquired on the whole sample are presented. Their positions are both dependent on strain and doping (strain and doping axis are represented by non-orthogonal vectors in this space). The spatially resolved doping (b) shows that region that are tightly coupled to h-BN present less electron doping. The colorscale in (a) represents the relative electron doping as in (b).}
  \label{doping}
\end{figure}

In Figure~\ref{doping}, we present a doping analysis of the sample realized prior to CL. We used Raman spectroscopy and tracked local shifts in the positions of the E$^{'}$ and A$^{'}_{1}$ modes (see the distribution in Figure~\ref{doping}a) to retrieve the spatial dependence of doping (Figure~\ref{doping}b). This is possible by discriminating the effect of strain, which alters the energy of the two Raman modes in a different way doping does. This kind of analysis was already outlined in the literature on monolayer MoS$_2$ \cite{michail2016,dubey2017}. We observe that regions exhibiting CL have a smaller $n$ doping, by typically several 10$^{12}$ cm$^{-2}$, compared to regions presenting no CL. 
Noting that the analysis concerns measurements acquired before the CL measurements, \textit{i.e.} prior to electron beam irradiation, we conclude that regions tightly coupled to h-BN are less electron doped than regions with a loose coupling. While the presence of an h-BN susbtrate has been shown to reduce electron doping compared to SiO$_2$ \cite{dubey2017}, in which case doping is created by charge traps at the interface \cite{Lu2014}, the exact mechanism at stake here to explain the coupling-dependent doping is more complex. Species trapped at the interface and gathered in the form of blisters are likely candidates to explain the observed $n$ doping. But the effect of, for instance, oxygen, that was present during the sample preparation in ambient atmosphere is still debated and might depend on the substrate \cite{Qi2016}. While the identification of the exact mechanism responsible for the observed reduction of $n$ doping in tightly coupled h-BN/MoS$_2$ regions will require further investigations, it is nevertheless possible to identify those regions using Raman spectroscopy as shown in Figure~\ref{doping}b. 

We note that we could not observe low-energy interlayer vibrational modes in Raman spectroscopy, that could be another signature, besides electronic doping level, of an intimate contact between MoS$_2$ and h-BN. Although such modes have indeed been observed in between TMDC layers\cite{Zhang2013,o'brien2016,liang2017}, we are not aware of any reports of such modes between h-BN and MoS$_2$.

\subsection{Cause of defect creation by electron-beam irradiation}
Having established that defects are induced by irradiation by the electron-beam used during the cathodoluminescence measurements, and that these defects yield prominently to non-radiative recombination of EXs, we now discuss their possible origin. We stress that all the signatures of the defects that we have presented here differentiate them from the ones that can be induced in h-BN upon electron irradiation  to create single photon emitters \cite{tran2016}. Besides being created at much lower energy (1~keV here vs 15~keV in Ref.~\cite{tran2016}), they do not show any optical activity and present a Raman signature.
Regarding their nature, the energy of the electron beam is well below the threshold for knock-on displacement of individual S or Mo atoms in MoS$_2$, which is several 10~keV and few 100~keV respectively \cite{Zan2013,AlgaraSiller2013,Garcia2014}. This effect can hence be ruled out as a source of defects in MoS$_2$ here. The dose in a typical CL experiment is estimated to be of the order of tens of mC/cm$^{2}$ (tens of electrons/nm$^{2 }$) which is several order of magnitude below the dose for typical transmission electron microscopy experiments \cite{Parkin2016}.

Another possible origin, also related to the electron beam, is a chemical reaction rather than a scattering effect. MoS$_2$ has indeed been shown to act as an active catalyst for hydrogen evolution reactions (HER). Following a Volmer-Heyrovsky type of mechanism, the electron beam used in our experiment might thus promote a multi-step chemical process proceeding for instance through the formation of MoH adducts \cite{Liao2013,Xie2013}. In our case, it is reasonable to assume that the blisters located at the MoS$_2$/h-BN interface contain airborne species such as water and/or oxygen that naturally adsorb on surfaces in ambient pressure conditions and will be trapped during the assembly of the heterostructure. It has been shown that the presence of oxygen can enhance catalytic reactions in MoS$_2$ \cite{Parzinger2015}. The trapped blisters may hence behave as aqueous solution micro- or nano-reactors. The chemical environment of the MoS$_2$ atoms bonded to hydrogen is different from that of the pristine material, which may allow the electron scattering processes needed to activate the above-discussed defect-induced Raman signatures. The exact nature of the defects is still an open question at this point and will require further studies. We note that the buried character of the interface makes traditional chemically sensitive surface probes (X-ray photoemission spectropscopy, local microscopy) not suited to such investigations. 

\subsection{Cathodoluminescence with improved spatial inhomogeneity}
The nanometer-scale spatial resolution of cathodoluminesence, as implemented in a scanning electron microscope, together with the tendency for defect formation under electron-beam irradiation of the blisters, represents a high resolution probe of the quality of the MoS$_2$/h-BN contact. Employing this probe allowed us to conclude that the PDMS stamping technique does not yield extended clean contacts beyond a few $\mu$m.

\begin{figure}
\includegraphics{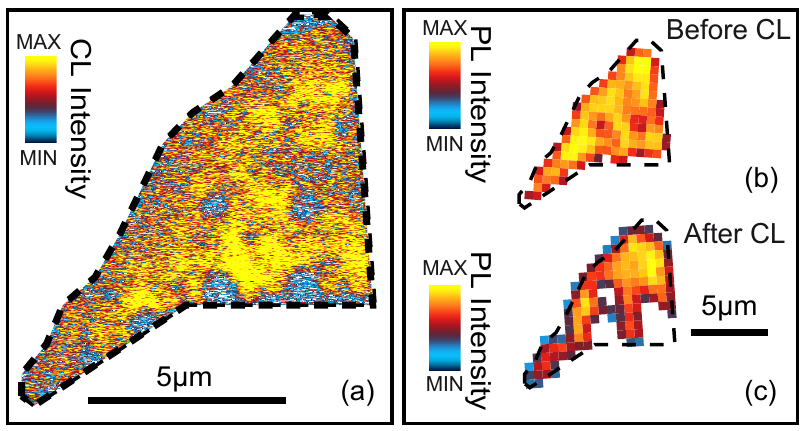}
  \caption{\textbf{Uniform coupling and properties using PPC pick-up preparation technique}. The CL signal measured at 5~K in the sample prepared by the pick-up technique is uniform across the flake. We observe sub-micrometric regions with reduced CL signal (blue) that are (air) bubbles. The integrated PL intensity measured at room temperature before (b) and after (c) CL shows a limited evolution compared to the sample made using PDMS. The absence of signal is limited to small regions and is caused by (air) bubbles created in the fabrication process.}
  \label{PPC}
\end{figure}

We then used an alternative transfer technique based on pick-up and drop-down with a PPC stamp. This technique allows to reduce the amount of blisters trapped at the interface \cite{Pizzocchero2016}. Figure~\ref{sample}e shows an optical micrograph of such a h-BN/MoS$_2$/h-BN heterostructure. The CL map of this sample is much more uniform than that of heterostructures prepared with a PDMS stamping as expected from the more uniform contact (compare Figures~\ref{CL}c and \ref{PPC}a).  Also PL measured before and after CL is very similar (see Figures~\ref{PPC}b,c) in contrast to the sample prepared using PDMS (see Figures~\ref{CL}d,e). It shows that electron beam irradiation in CL has not created extended defective regions because of the uniform coupling in that sample. The complete suppression of bubbles should allow further optimization of the process \cite{Pizzocchero2016}.

\section{Conclusions}
Our work shows that clean interfaces between TMDCs (here MoS$_2$) and h-BN are required to allow efficient charge transfer between the barrier and active material. Assembly techniques that are commonly employed to prepare heterostructures often trap blisters of contaminants at the interface between the TMDC and h-BN surfaces. Contrary to what may have been thought, the detrimental effect of such blisters is not only a hindrance of charge transfers at the interface. Cathodoluminescence is there quenched due also to electron-beam-induced damage of the TMDC crystal. Defects are generated, and induce non-radiative charge carrier recombination. We ascribe the formation of defects to an electron-promoted chemical reaction. We find that the cleanliness of the interface is of superior spatial uniformity, and the generation of defects is greatly avoided, when a pick-up/drop-down assembly technique with PPC is employed.

The narrow emission linewidth observed in CL and the localized electron beam should allow to spatially map strain and doping profiles with nanometer resolution by analyzing the exciton peak position and the presence of trion emission. CL could also be used to study with unprecedented spatial resolution single photon emitters that were reported in several TMDCs \cite{Tonndorf2015,He2015,Koperski2015,Srivastava2015,Kumar2015}.

\begin{acknowledgments}
This work was supported by the French National Research Agency (ANR) in the framework of the J2D project (ANR-15-CE24-0017), the 2DTransformers project under OH-RISQUE program (ANR-14-OHRI-0004), and of the "Investissements d’avenir" program (ANR-15-IDEX-02). J.R. acknowledges support from Grenoble Alpes University community (AGIR-2016-SUGRAF). G.N., A. B. and V.B. thank support from CEFIPRA. Growth of hexagonal boron nitride crystals was supported by the Elemental Strategy Initiative conducted by the MEXT, Japan and the CREST (JPMJCR15F3), JST. We thank the Nanofab group at Institut Néel for help with van der Waals heterostructures preparation setup. We thank C. Bucher for fruitful discussions.
\end{acknowledgments}

\bibliography{Nayak_CL_MoS2_bib}

\end{document}